\documentclass[prl,twocolumn,letterpaper,showpacs,groupedaddress,superscriptaddress,floatfix,preprintnumbers,tightenlines]{revtex4-1}

\usepackage[colorlinks,urlcolor=blue]{hyperref}
\usepackage{amsmath,amssymb,amsfonts}
\usepackage{graphicx}
\usepackage{bm}
\usepackage{comment}
\usepackage{color}
\usepackage{csquotes}
\usepackage{qcircuit}
\usepackage[section]{placeins}
\usepackage[table]{xcolor}
\usepackage{balance}
\usepackage[utf8]{inputenc}

\allowdisplaybreaks
\let\Oldsection\section
\renewcommand{\section}{\FloatBarrier\Oldsection}
\let\Oldsubsection\subsection
\renewcommand{\subsection}{\FloatBarrier\Oldsubsection}

\newcount\hour \newcount\hourminute \newcount\minute
\hour=\time \divide \hour by 60
\hourminute=\hour \multiply \hourminute by 60
\minute=\time \advance \minute by -\hourminute
\newcommand{\mydate}{\ \today \ - \number\hour :\number\minute}

\def\OMIT#1{{}}

\newcommand\scalemath[2]{\scalebox{#1}{\mbox{\ensuremath{\displaystyle #2}}}}
\begin{document}

\title{ SU(2) non-Abelian gauge field theory in one dimension on digital quantum computers }

\author{Natalie Klco, Jesse R.~Stryker and Martin J.~Savage}
\affiliation{Institute for Nuclear Theory, University of Washington, Seattle, WA 98195-1550, USA}

\date{\mydate}

\preprint{INT-PUB-19-033}


\begin{abstract}
\noindent
An improved mapping of one-dimensional SU(2) non-Abelian gauge theory onto qubit degrees of freedom is presented.
This new mapping allows for a reduced unphysical Hilbert space. Insensitivity to interactions within this unphysical space is exploited to design more efficient quantum circuits.
Local gauge symmetry is used to analytically incorporate the angular momentum alignment, leading to qubit registers encoding the total angular momentum on each link.  The results of a multi-plaquette calculation on IBM's quantum hardware are presented.
\end{abstract}
\maketitle

Non-Abelian gauge field theories play a central role in the description of the known forces of nature.
Since the early 1970's, the strong interactions that define the nuclear forces and the dynamics of quarks and gluons in the early universe
are known to emerge from an unbroken SU(3) local gauge symmetry,
defining quantum chromodynamics (QCD)~\cite{Politzer:1973fx,Gross:1973id,Politzer:1974fr}.
Similarly, the electroweak interactions are known to result from the spontaneous breaking of SU(2)$_L\otimes$U(1)$_Y$ local
gauge symmetries~\cite{Weinberg:1967tq,Glashow:1961tr,Salam:1968rm,Higgs:1964pj}.
Great success has been achieved in computing the properties and low-energy dynamics of
hadronic systems using the numerical technique of lattice QCD~\cite{Wilson:1974sk,Creutz:1980zw} on the world's largest supercomputers.
Current lattice QCD calculations at the physical quark masses have resulted from a sustained co-development
effort over the last $\sim50$ years.
Those developments began with calculations on small lattices, with unphysical quark masses, and with large lattice spacings using computers available
during the 1970's~\cite{Creutz:1980zw}.
While good progress is being made in designing~\cite{Jordan:2011ne,Jordan:2011ci,Wiese:2013uua,Wiese:2014rla,Jordan:2014tma,Garcia-Alvarez:2014uda,Zohar:2015hwa,Pichler:2015yqa,Bazavov:2015kka,Mezzacapo:2015bra,Dalmonte:2016alw,Somma:2016:QSO:3179430.3179434,Bermudez:2017yrq,Macridin:2018gdw,Zache:2018jbt,Preskill:2018fag,Zhang:2018ufj,Kaplan:2018vnj,Klco:2018zqz,Raychowdhury:2018osk,Gustafson:2019mpk} and implementing~\cite{Zohar:2012ay,Zohar:2012xf,Banerjee:2012pg,Banerjee:2012xg,Marcos:2014lda,Marshall:2015mna,Zohar:2016iic,Martinez2016,Klco:2018kyo,Lu:2018pjk,Klco:2019xro,Yeter-Aydeniz:2018mix,Kokail:2018eiw,Davoudi:2019bhy} quantum field theories for quantum devices,
non-Abelian gauge theories have not yet been simulated on today's limited and noisy hardware.
It is in the spirit of the early days of lattice gauge theory that we develop an improved algorithm to evolve a string of SU(2) plaquettes, and use it to simulate a non-Abelian gauge field theory on IBM's digital quantum hardware.

The Hamiltonian formulation of lattice gauge theories \cite{KogutSusskind1975} includes exponentially-large sectors of unphysical~\footnote{The space referred to as unphysical can be naturally interpreted as isolated Hilbert spaces with non-zero external sources.} Hilbert space in order to maintain spatially-local interactions while satisfying gauge constraints.
The hardware error rates and gate fidelities of current NISQ-era~\cite{Preskill2018quantumcomputingin} quantum devices, and the lack of error correction capabilities, allow quantum states to  disperse into these unphysical sectors.
To avoid such dispersion, previous quantum simulations of lattice gauge theories have employed various procedures to remove the unphysical Hilbert space from the embedding onto quantum devices~\cite{Martinez2016,Muschik:2016tws,Klco:2018kyo,Kokail:2018eiw,Lu:2018pjk}.
However, these techniques do not scale efficiently, and a generic description for multi-dimensional lattices with non-trivial gauge groups in terms of only local, physical degrees of freedom is not currently known.
A variety of approaches to simulating gauge theories are being pursued---reformulating the interactions, lattice structure, and degrees of freedom by designing Hilbert space bases of group elements, Schwinger bosons, duality transformations, loop variables, tensor networks, and more~\cite{KogutSusskind1975,PhysRevD.31.3201,Burgio:1999tg,Mathur_2005,Mathur:2005fb,PhysRevA.73.022328,Anishetty_2009,Zohar:2012xf,Orland:2013faa,Stannigel:2013zka,Rizzi:2014bba,Anishetty:2014tta,Zohar:2014qma,Kuno:2014npa,Liu:2014uhw,Hu:2014fna,Tagliacozzo:2014bta,Bazavov:2015kka,Zohar:2016iic,Kuno:2016ipi,Zohar:2018cwb,Bender:2018rdp,Hackett:2018cel,Kaplan:2018vnj,Dreher:2018qls,Stryker:2018efp,Raychowdhury:2018osk,Li:2018vjg,Zohar:2018nvl,Lamm:2019bik,Zohar:2019ygc}---often with the explicit goal of mitigating unphysical degrees of freedom.
Reductions have been obtained by solving Gauss's law,
which is related to loop formulations in which the fundamental degrees of freedom are gauge invariant
\cite{Polyakov:1979gp,Mandelstam:1978ed,Nambu:1978bd,Polyakov:1980ca,Furmanski:1986bj,Gambini:1988kq,dibartolo.gambini.eaHamiltonianLattice89,Anishetty:1990en,Bruegmann:1991wg,Watson:1993zr,Mathur:2005fb,Anishetty:2014tta,Raychowdhury:2018tfj}.
Proposed for both analog and digital quantum implementation, progress is being made toward using renormalization group methods to connect quantum link models~\cite{HORN1981149,ORLAND1990647,Chandrasekharan:1996ih,Brower:1997ha,Wiese:2014rla,Tagliacozzo:2012df,Banerjee:2012xg,Mezzacapo:2015bra}
to continuum theories of importance~\cite{Brower:1997ha,Meurice:2016mkb,BrowerAspen2019,MeuriceAspen2019}.

In this work, the angular momentum basis~\cite{KogutSusskind1975,PhysRevD.31.3201,Burgio:1999tg} is utilized,
made computationally feasible on quantum devices by
exploiting the local gauge symmetry to remove the angular momentum alignment variables.
Time evolution of a one-dimensional string of two SU(2) plaquettes is then implemented on IBM's {\tt Tokyo}~\cite{ibm} quantum device with employed error mitigation techniques.
This geometry, involving only three-point vertices, enables the exploration of SU(2) gauge field interactions with qubit requirements that depend linearly on the spatial volume.

\begin{figure}
\includegraphics[width=\columnwidth]{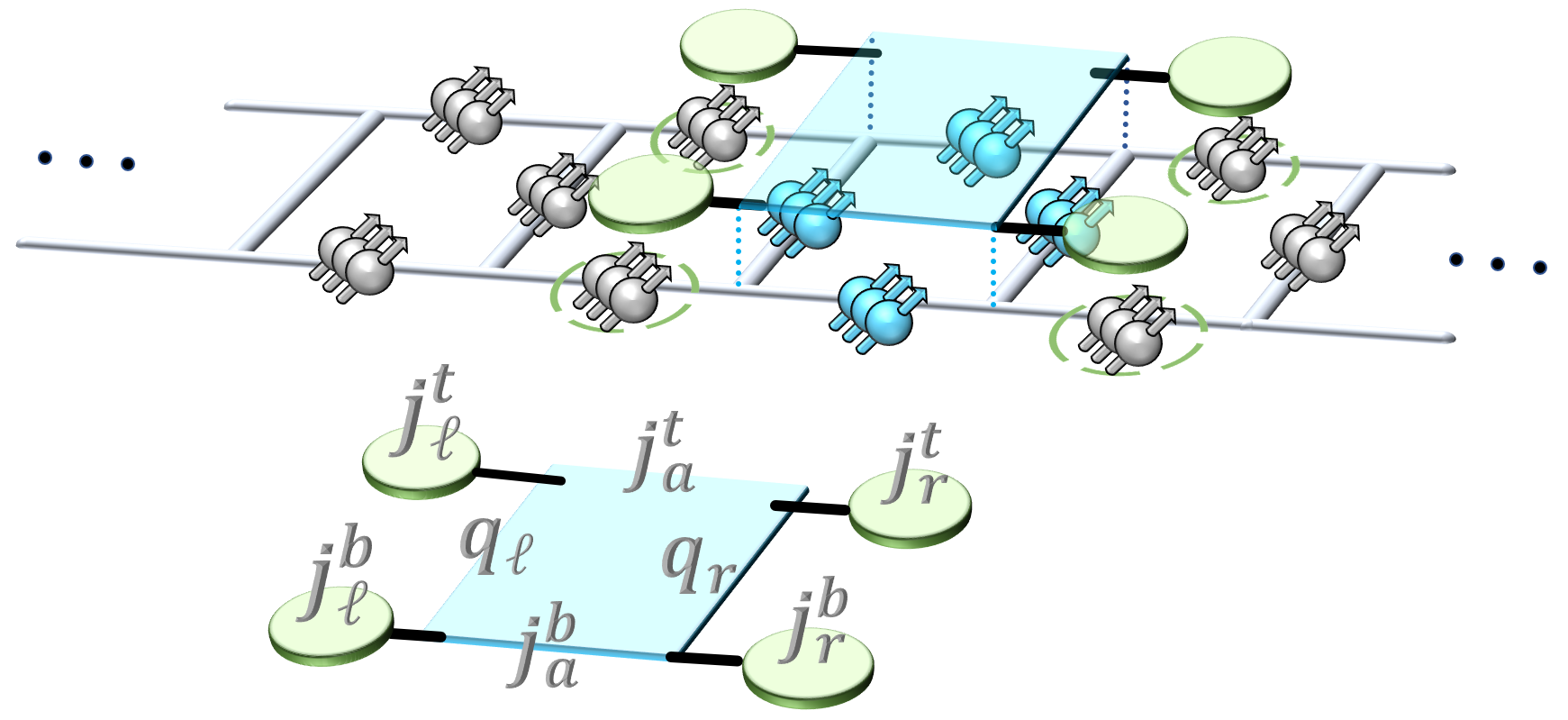}
\caption{(top)
Diagram of the lattice distribution of $\lceil\log_2(2\Lambda_j+1)\rceil$-qubit registers and the action of
$\hat{\Box}$ on SU(2) plaquettes in one dimension.
$\hat{\Box}$ operates on the four qubit registers in the plaquette
and is controlled by the four neighboring qubit registers to enforce the Gauss's law constraint. (bottom)
The plaquette operator with labeled angular momentum registers.
}
\label{fig:diagram}
\end{figure}

The Hamiltonian of spatially-discretized Yang-Mills gauge theory is~\cite{KogutSusskind1975} (in lattice units)
\begin{equation}
  \hat{H} = \frac{g^2}{2} \sum_{\text{links}} \hat{E}^2 - \frac{1}{2g^2} \sum_{\Box} \left( \hat{\Box} + \hat{\Box}^\dagger \right)
\end{equation}
where $\hat{E}^2$ is the local gauge-invariant Casimir operator, $\hat{\Box}$ is the gauge-invariant plaquette operator contracting closed loops of link operators, and $ \hat{\Box}  = \hat{\Box}^\dagger$ for SU(2). 
On a square lattice, the single plaquette operator is
\begin{equation}
  \hat{\Box} = \sum_{\alpha, \beta, \gamma, \delta
  =
  -\frac{1}{2}}^{\frac{1}{2}}
  \hat{U}_{\alpha \beta}\
  \hat{U}_{\beta \gamma}\
  \hat{U}_{\gamma \delta} \
  \hat{U}_{\delta \alpha} \
  \label{eq:BoxOperator}
\end{equation}
where $\hat{U}_{\alpha\beta}$ is a $j = 1/2$ link operator with definite starting and ending points oriented around a plaquette.
In the limit of strong coupling, $g^2\rightarrow\infty$,
 this Hamiltonian is dominated by the electric contributions
 and fluctuations between configurations of definite link angular momentum vanish.
In weak coupling, the magnetic contributions dominate and a theory of dynamical loops emerges.
The angular momentum basis describes the quantum state of a generic link by its irreducible representation, $j$,
and associated third-component projections at the left and right end of the link in the ${\bf 2}$ and $\bar{\bf 2}$ representations,
$|j, m, m'\rangle \equiv |j,m\rangle \otimes |j,m'\rangle $, respectively.
In one dimension, SU(2) lattice gauge theory can be spatially discretized onto
a string of plaquettes (see Fig.~\ref{fig:diagram}).
With periodic boundary conditions (PBCs), only three-point
vertices contribute to such a plaquette chain.
To form gauge singlets,
components of the three links at each vertex
are contracted with an SU(2) Clebsch-Gordan coefficient.
The wavefunction at each vertex has the form
\begin{equation}
\scalemath{0.9}{
V\  \sim\   \sum_{b,c,e} \langle j_1, b, j_2, e |q, c\rangle
\ |j_1, a, b\rangle \otimes |q, c, d\rangle \otimes |j_2, e, f\rangle \ ,  }
\end{equation}
where indices $b, c, $ and $e$ are located at the vertex.
By focusing on a system with an even number of plaquettes,
matrix elements of the arbitrary plaquette operator may be determined.
The state of an even-length lattice in one dimension with PBCs and with definite link angular momenta is
\begin{align}
  |\chi\rangle &= \mathcal{N} \sum_{\{m\} } \prod_{i = 1}^L \langle j_i^t, m_{i,R}^t, j_{i+1}^t, m_{i+1,L}^t| q_i, m_{q_i}^t \rangle \\
  & \hspace{2cm}
  \langle j_i^b, m_{i,R}^b, j_{i+1}^b, m_{i+1,L}^b| q_i, m_{q_i}^b \rangle  \nonumber \\ & \hspace{0.4cm}  |j_i^t, m_{i,L}^t, m_{i,R}^t\rangle \otimes |j_i^b, m_{i,L}^b, m_{i,R}^b \rangle \otimes  |q_i, m_{q_i}^t, m_{q_i}^b\rangle \nonumber
\end{align}
with $j_{L+1} = j_1$, $m_{L+1} = m_1$, and normalization $\mathcal{N}=\prod_i (\dim(q_i))^{-1}$
with $\dim(q) = 2q+1$.
Referring to the plaquette string's ladder structure, on links located as rungs of the ladder, angular momentum values are labeled by $q$. Thus, a plaquette string is created by two strings of $j$-type registers connected periodically by rungs of $q$-type registers.
The contraction with Clebsch-Gordan coefficients at each vertex ensures the local gauge singlet structure
required by Gauss's law.
The link operator acts on the degrees of freedom at each end of a link and is a source of $j = 1/2$ angular momentum,
\begin{multline}
  {\hat U}_{\alpha \beta} |j, a, b\rangle 
  = \sum_{\oplus J} \sqrt{\frac{\text{dim}(j)}{\text{dim}(J)}}
  \     |J, a+\alpha, b+\beta\rangle  \\*
 \times\
  \langle j, a, \frac{1}{2}, \alpha| J, a+\alpha  \rangle
 \langle j, b, \frac{1}{2}, \beta| J, b+\beta\rangle \ \ , \label{LinkOperator}
\end{multline}
which contains non-vanishing contributions only for $J = j\pm \frac{1}{2}$~\cite{Zohar:2014qma}.
It follows that matrix elements of the plaquette operator in one dimension are
\begin{align}
  \langle \chi_{\cdots, j^{t,b}_{\ell}, q_{\ell f}, j^{t,b}_{af}, q_{r f}, j^{t,b}_{r}, \cdots } | \hat{\Box} | \chi_{\cdots, j^{t,b}_{\ell}, q_{\ell i}, j^{t,b}_{ai}, q_{r i}, j^{t,b}_{r}, \cdots } \rangle &= \nonumber\\* &\hspace{-6cm}
    \sqrt{\dim(j_{ai}^t)\dim(j_{af}^t) \dim(j_{ai}^b)\dim(j_{af}^b) } \nonumber \\*  &\hspace{-6.5cm} \times \sqrt{\dim(q_{\ell i})\dim(q_{\ell f}) \dim(q_{r i})\dim(q_{r f})} \label{eq:plaquettematrixelement} \\*  &\hspace{-6.5cm}
   \times (-1)^{j_\ell^t + j_\ell^b + j_r^t + j_r^b + 2(j_{af}^t + j_{af}^b -q_{\ell i} -q_{r i} )} \nonumber \\* & \hspace{-7.7cm}
  \times \scalemath{0.9}{\begin{Bmatrix}
    j_{\ell}^t & j_{ai}^t & q_{\ell i} \\
    \frac{1}{2} & q_{\ell f} & j_{af}^t
  \end{Bmatrix} \begin{Bmatrix}
    j_{\ell}^b & j_{ai}^b & q_{\ell i} \\
    \frac{1}{2} & q_{\ell f} & j_{af}^b
  \end{Bmatrix} \begin{Bmatrix}
    j_{r}^t & j_{a i}^t & q_{ri} \\
    \frac{1}{2} & q_{rf} & j_{af}^t
  \end{Bmatrix} \begin{Bmatrix}
    j_{r}^b & j_{ai}^b & q_{ri} \\
    \frac{1}{2} & q_{rf} & j_{af}^b
  \end{Bmatrix}} \nonumber
\end{align}
where the indices $j^{t,b}_\ell, q_{\ell i } ,q_{\ell f } , j^{t,b}_a, q_{r i}, q_{r f},$ and $j^{t,b}_r$ are used to indicate the $j$-values relevant for the single plaquette operator
(as depicted in Fig.~\ref{fig:diagram}) and the brackets indicate Wigner's 6-j symbols.
The four registers $j^{t,b}_{\ell,r}$ outside the plaquette are not modified by the action of the plaquette operator.
However, their inclusion as control registers is necessary to maintain Gauss's law.
The sums over alignment in each gauge-invariant space yield a dramatically reduced
Hilbert space to be mapped onto a quantum device,
characterized entirely by the $|j\rangle$'s (rather than the $|j, m, m'\rangle$'s) incrementing naturally by half-integers.
The qubit representation of the periodic plaquette string is shown on the top panel of Fig.~\ref{fig:diagram}, where each link contains a $\lceil \log_2 (2 \Lambda_j +1) \rceil$-qubit register with $\Lambda_j$ the angular momentum truncation per link.

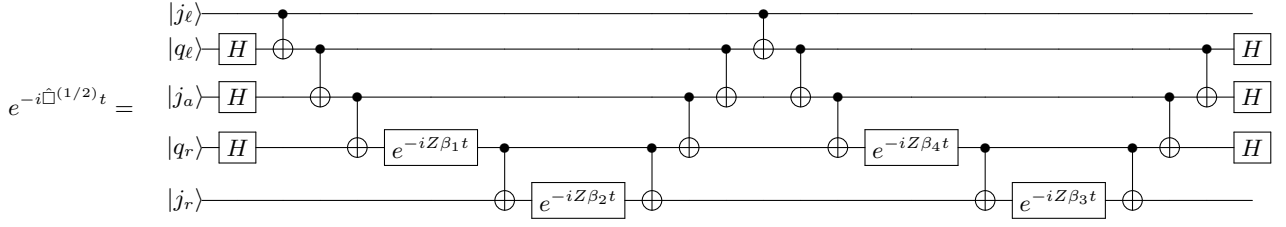
\begin{figure*}[!ht]
  \begin{equation}
\scalemath{1.0}{
e^{-i \hat{\Box}^{(1/2)}t} = \qquad \
\begin{gathered}
  \Qcircuit @R 0.7em @C 0.7em {
    |j_\ell\rangle \quad \ &\qw &\ctrl{1} &\qw &\qw &\qw &\qw &\qw &\qw &\qw &\qw &\ctrl{1}&\qw &\qw &\qw &\qw &\qw &\qw &\qw &\qw &\qw \\
|q_\ell\rangle \quad \ &\gate{H}&\targ&\ctrl{1}&\qw &\qw&\qw&\qw&\qw&\qw&\ctrl{1}&\targ&\ctrl{1}&\qw &\qw&\qw&\qw&\qw&\qw&\ctrl{1}&\gate{H} \\
|j_a\rangle \quad \ &\gate{H}&\qw&\targ&\ctrl{1}&\qw&\qw&\qw&\qw&\ctrl{1}&\targ&\qw&\targ&\ctrl{1}&\qw&\qw&\qw&\qw&\ctrl{1}&\targ&\gate{H} \\
|q_r\rangle \quad \ &\gate{H}&\qw&\qw &\targ&\gate{e^{-i Z \beta_1t}}&\ctrl{1}&\qw&\ctrl{1}&\targ&\qw&\qw&\qw&\targ&\gate{e^{-i Z \beta_4t}}&\ctrl{1}&\qw&\ctrl{1}&\targ&\qw&\gate{H} \\
|j_r\rangle \quad \ &\qw &\qw&\qw &\qw &\qw&\targ&\gate{e^{-i Z \beta_2t}}&\targ&\qw&\qw&\qw&\qw&\qw &\qw&\targ&\gate{e^{-i Z \beta_3t}}&\targ&\qw&\qw&\qw \\
  }
\end{gathered}
  }\nonumber
\end{equation}
\caption{
Digital circuit implementation of the plaquette operator
centered on $j_a$ for a truncated lattice with $\Lambda_j = 1/2$.
}
\label{fig:circuitNp}
\end{figure*}
\begin{figure*}[!ht]
\centering
\begin{equation}
\scalemath{1.0}{
e^{-i \hat{\Box}_2^{(1/2)}t} = \quad
  \begin{gathered}
    \Qcircuit @R 0.7em @C 0.7em {
    |j_{\ell}\rangle\quad&\ctrlo{1} & \ctrl{1} & \qw \\
    |q_\ell\rangle\quad&\multigate{2}{e^{-i XXX t}} & \multigate{2}{e^{-i \frac{1}{4}  XXX t}}& \qw \\
    |j_a\rangle\quad& \ghost{e^{-i XXX t}} & \ghost{e^{-i \frac{1}{4}  XXX t}} &\qw \\
    |q_r\rangle\quad& \ghost{e^{-i  XXX t}} & \ghost{e^{-i   \frac{1}{4}  XXX t}} &\qw \\
    }
  \end{gathered} =
  \begin{gathered}
    \Qcircuit @R 0.7em @C 0.7em {
    & \qw & \qw & \qw & \qw & \targ & \gate{e^{-i \tilde{\beta}_1 Z t}} & \targ & \qw & \qw & \qw \\
    & \gate{H} &  \qw & \targ & \gate{e^{-i \tilde{\beta}_2 Z t }} & \ctrl{-1} & \qw & \ctrl{-1} & \targ & \qw & \gate{H} \\
    & \gate{H} & \targ & \ctrl{-1} & \qw & \qw & \qw & \qw & \ctrl{-1} & \targ & \gate{H} \\
    & \gate{H} & \ctrl{-1} & \qw & \qw & \qw & \qw & \qw & \qw & \ctrl{-1} & \gate{H}\\
    }
  \end{gathered} \ \ \ \
  \begin{gathered}
  \includegraphics[width = 0.2\textwidth]{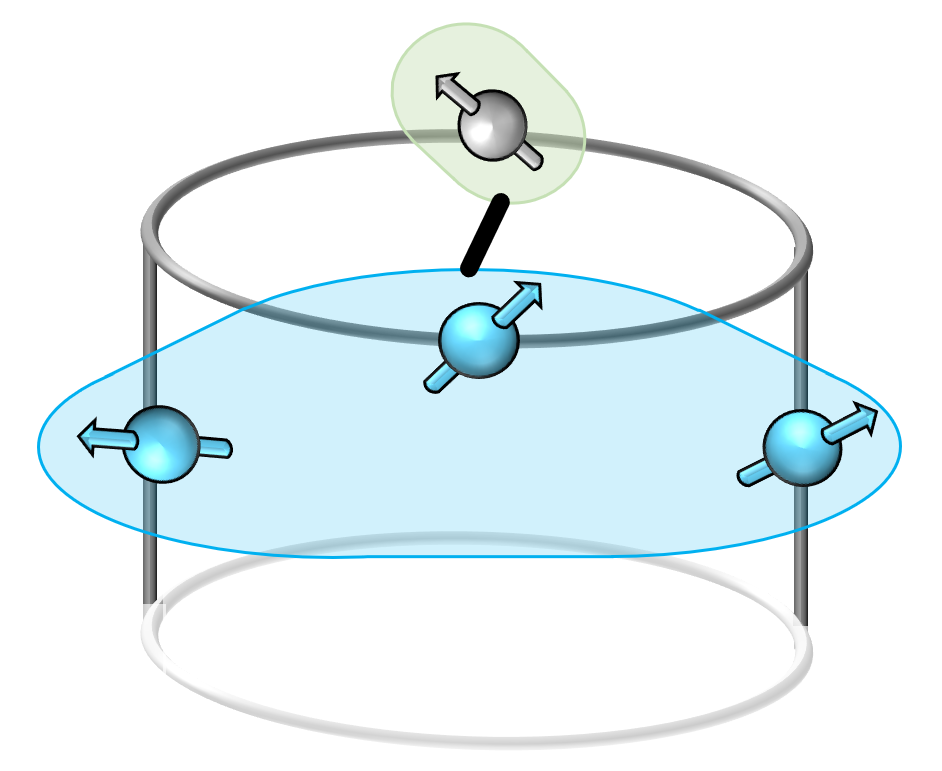}
  \end{gathered}
  \nonumber
  }
\end{equation}
\caption{
Digital circuit implementation of the plaquette operator centered on $j_a$ for a truncated lattice with $\Lambda_j = 1/2$, two plaquettes, and PBCs as depicted at the right.
}
\label{fig:circuit2p}
\end{figure*}

In the following, circuits are devised for the plaquette operator with
angular momentum truncation $\Lambda_j = 1/2$.
For time evolution beginning in the strong-coupling vacuum,
the top and bottom $j$ values are equivalent with this cutoff due to SU(2) flux conservation.
As a result,
the bottom $j$ registers need not be mapped onto the quantum
device~\footnote{
A similar argument shows that one $q$-type link register could be removed from the qubit mapping at the
expense of incorporating non-local operators.
} and the plaquette operator reduces to a five-qubit operator.
\begin{table}[!ht]
{\renewcommand{\arraystretch}{1.4}
  \begin{tabular}{cc}
  \hline
  \hline
    $\langle j_{\ell f}\  q_{\ell f}\  j_{af}\ q_{rf}\ j_{rf}| \hat{\Box}^{(1/2)} |j_{\ell i}\ q_{\ell i}\ j_{ai}\ q_{ri}\ j_{ri}\rangle$ & \\
  \hline
  \hline
    $\langle 00000 | \hat{\Box}^{(1/2)} | 0 \frac{1}{2} \frac{1}{2} \frac{1}{2} 0\rangle, \langle 0 \frac{1}{2} \frac{1}{2} \frac{1}{2} 0 | \hat{\Box}^{(1/2)} | 00000\rangle$ & 1 \\
    $\langle 000\frac{1}{2} \frac{1}{2} | \hat{\Box}^{(1/2)} | 0 \frac{1}{2}  \frac{1}{2} 0 \frac{1}{2} \rangle, \langle 0 \frac{1}{2}  \frac{1}{2} 0 \frac{1}{2}  | \hat{\Box}^{(1/2)} | 000\frac{1}{2} \frac{1}{2} \rangle$ & $\frac{1}{2}$ \\
    $\langle \frac{1}{2} \frac{1}{2} 000| \hat{\Box}^{(1/2)} | \frac{1}{2} 0 \frac{1}{2}\frac{1}{2}  0 \rangle, \langle \frac{1}{2} 0 \frac{1}{2}\frac{1}{2}  0| \hat{\Box}^{(1/2)} | \frac{1}{2} \frac{1}{2} 000  \rangle$ & $\frac{1}{2}$ \\
    $\langle \frac{1}{2}  0 \frac{1}{2}  0 \frac{1}{2}  | \hat{\Box}^{(1/2)} | \frac{1}{2}\frac{1}{2}  0 \frac{1}{2} \frac{1}{2} \rangle, \langle \frac{1}{2}\frac{1}{2}  0 \frac{1}{2} \frac{1}{2}  | \hat{\Box}^{(1/2)} | \frac{1}{2}  0 \frac{1}{2}  0 \frac{1}{2}  \rangle$ & $\frac{1}{4}$ \\
  \hline
  \hline
  \end{tabular}
  }
  \caption{
  Non-zero matrix elements of the $\Lambda_j = 1/2$ plaquette operator $\hat{\Box}^{(1/2)}$ as calculated in Eq.~\eqref{eq:plaquettematrixelement} with $j_{\ell,a,r}^t = j_{\ell,a,r}^b$.  All other matrix elements between physical states are zero.
  }
  \label{tab:matrixelements}
\end{table}

While matrix elements of the plaquette operator in the physical space are critical,
those in the unphysical space are not.
As long as the matrix elements mixing the two spaces vanish,
there exists significant freedom in designing the structure of the unphysical space
to hardware-specifically optimize quantum computation.
For example, here
it is  convenient to identify a
\emph{gauge-variant completion} (GVC)
within the set of Pauli operators to minimize the quantum gate resource requirements.
Observing the plaquette operator matrix elements in Table~\ref{tab:matrixelements},
states are connected when $q_\ell, j_a, $ and $q_r$ experience a qubit inversion with a
matrix element dependent on the $j_\ell,j_r$-sector.
Such a controlled operator is depicted schematically at the bottom of
Fig.~\ref{fig:diagram} (with top and bottom $j$'s identified) and may be written as
\begin{align}
  \hat{\Box}^{(1/2)} &= \Pi_0 XXX \Pi_0 + \frac{1}{2} \Pi_0 XXX \Pi_1 \label{eq:boxprojectors}\\ & \hspace{1cm}+ \frac{1}{2} \Pi_1 XXX \Pi_0 + \frac{1}{4} \Pi_1 XXX \Pi_1 \nonumber
\end{align}
with $\Pi_0 = \frac{1}{2} (\mathbb{I} + Z) $ and $\Pi_1 = \frac{1}{2} ( \mathbb{I} - Z)$,
the $j = 0(\frac{1}{2})$ state mapped to quantum state $|0\rangle(|1\rangle)$,
and the Hilbert spaces ordered as in the heading of Table~\ref{tab:matrixelements}.
With this GVC, the plaquette Hamiltonian has 24 non-zero couplings between unphysical states.
One possible digital qubit implementation of the associated time evolution operator is shown explicitly in Fig.~\ref{fig:circuitNp} with gate rotations defined by linear combinations of $\hat{\Box}^{(1/2)}$ matrix elements, as established in Ref.~\cite{Klco:2019xro}, described by the following matrix structure:
\begin{equation}
  \vec{\beta} =\begin{pmatrix}
    1 & 1 & 1 & 1 \\
    1 & -1 & -1 & 1 \\
    -1 & -1 & 1 & 1 \\
    -1 & 1 & -1 & 1 \\
  \end{pmatrix}^{-1} \begin{pmatrix}
    1 \\ 1/2 \\ 1/2 \\ 1/4
  \end{pmatrix} \ \ \ .
  \label{eq:betaLC}
\end{equation}
As written, this operator acts equivalently throughout the one-dimensional string of plaquettes.


Specializing to the two-plaquette system with PBCs, only the matrix elements in the first and last rows of
Table~\ref{tab:matrixelements} remain.
The second plaquette operator in the two-plaquette system reduces to the following four-qubit operator,
\begin{equation}
  \hat{\Box}^{(1/2)}_{2} = \Pi_0 XXX + \frac{1}{4} \Pi_1 XXX \ \ \ .
  \label{eq:plaquette2fourq}
\end{equation}
Digital implementation of this operator is shown in Fig.~\ref{fig:circuit2p} with the reduced linear combination structure defined by the first and fourth rows and columns of the matrix shown in Eq.~\eqref{eq:betaLC} giving rise to the vector $\widetilde{\beta}$.
A natural qubit representation of the electric operator is
\begin{equation}
  \hat{H}_E^{( 1/2
  )} = \frac{g^2}{2} \sum_{\text{links}} \frac{3}{4} \left( \frac{\mathbb{I}-Z}{2} \right) \ \ \ ,
  \label{eq:electricHamiltonian2P}
\end{equation}
including 12 non-zero elements in the unphysical Hilbert space.


%
\begin{figure}
  \centering
  \includegraphics[width = 0.9\columnwidth]{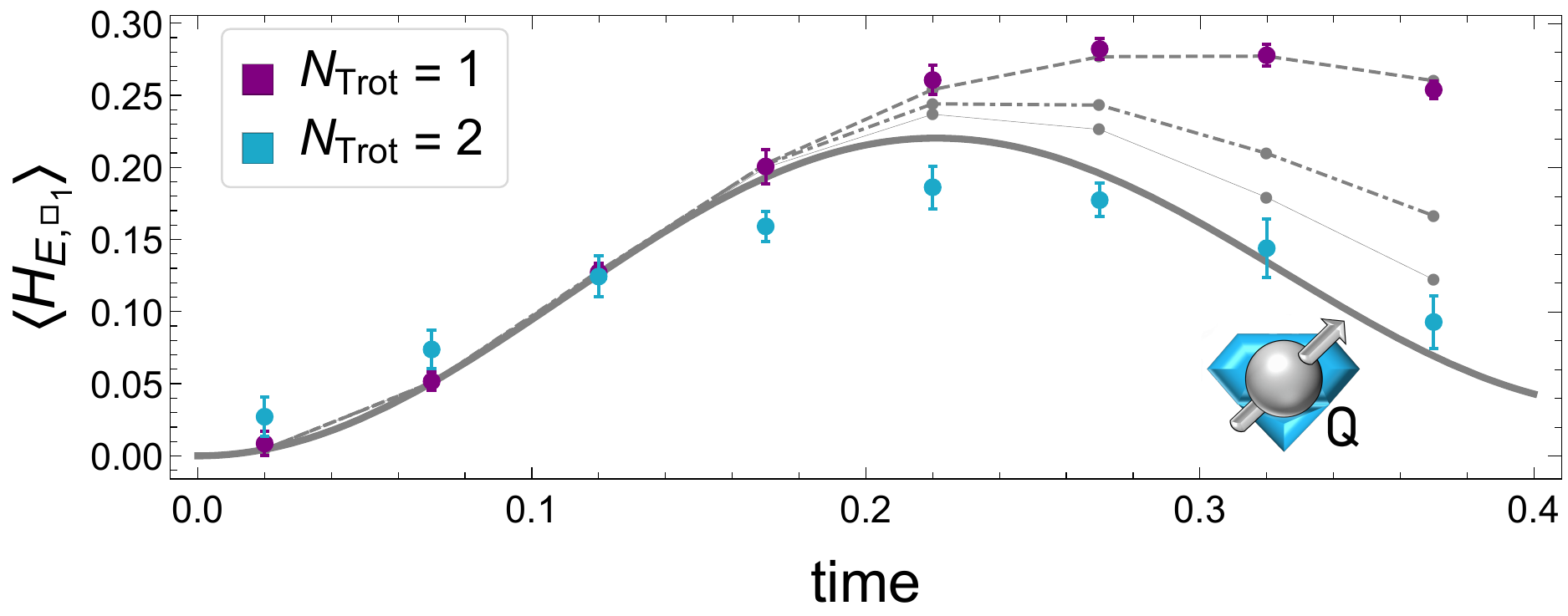}
  \includegraphics[width=0.9\columnwidth]{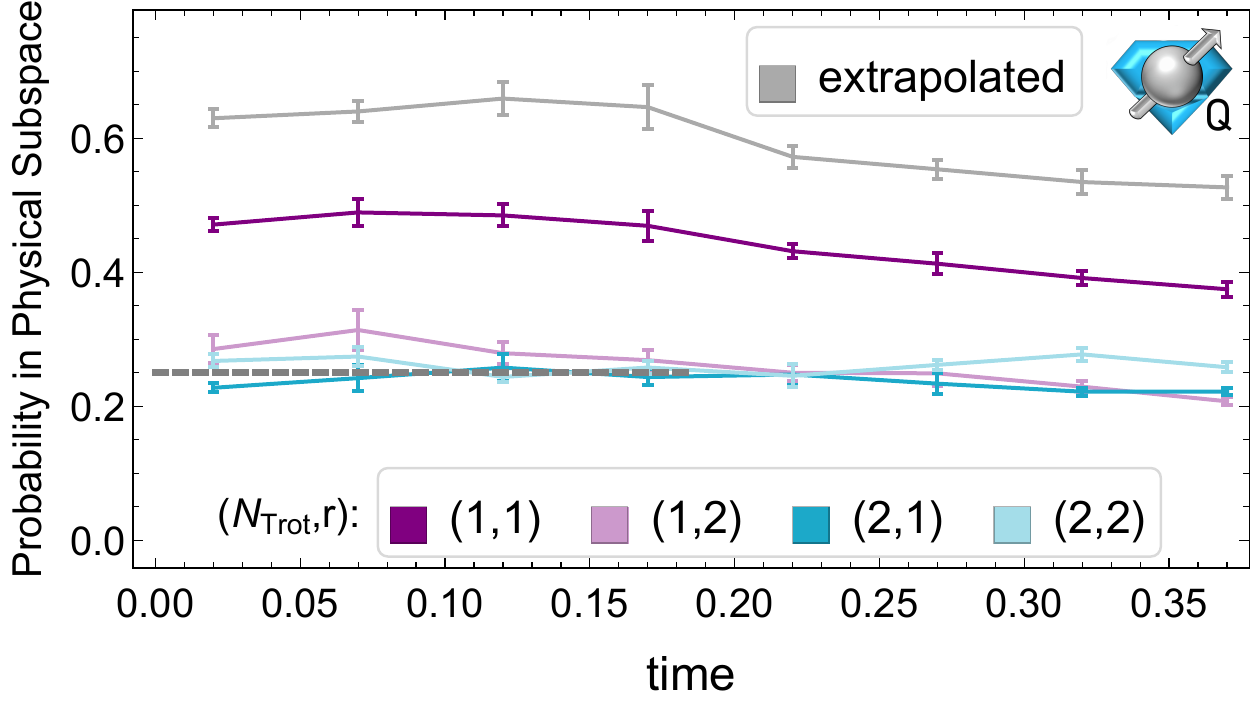}
  \caption{
(top) Expectation value of the electric energy contribution of the first plaquette in the two-plaquette lattice with PBCs and coupling $g^2 = 0.2$ computed on IBM's {\tt Tokyo}.  The dashed, dot-dashed, and thin gray lines are the $N_{\text{Trot}} = 1, 2, 3$ Trotterized expectation values, while the thick gray line is the exact time evolution.   (bottom) Measured survival probability to remain in the physical subspace.
Uncertainties represent statistical variation, as well as a systematic uncertainty estimated from reproducibility measurements.
The icons appearing are defined in Ref.~\cite{Klco:2019xro}.
  }
  \label{fig:IBMdata}
\end{figure}
Real-time evolution of two plaquettes with PBCs (see the right panel of Fig.~\ref{fig:circuit2p})
and truncation $\Lambda_j= 1/2$ has been implemented on IBM's quantum device {\tt Tokyo},
selected for its available connectivity of a four-qubit loop.
The top panel of Fig.~\ref{fig:IBMdata} shows time-evolved expectation values of the energy contributions from the first electric plaquette calculated with one and two Trotter steps~\footnote{The Trotter step in this calculation has been ordered in application as the first plaquette, the second plaquette as written in Eq.~\eqref{eq:plaquette2fourq}, and lastly the electric time evolution operator.}.
The electric contributions, being localized in their measurement to the four-dimensional physical subspace, are well determined after a small number of samples.
In contrast, measuring the energy contributions from the magnetic Hamiltonian requires increased sampling due to the operator's natural representation in the Pauli-$X$ basis of the $q_\ell, j_a$, and $q_r$ qubit registers---distributing the wavefunction's amplitude throughout the Hilbert space.
Results have been corrected for measurement error by the constrained inversion of a 16-dimensional calibration
matrix informed by preparation of each of the 16 computational basis states prior to calculation.
The resulting probabilities are linearly extrapolated in the presence of CNOT gates, post-selected within the
gauge-invariant space, and renormalized.
The linear extrapolation is informed by $r = 1, 2$, where $r = 1$ is the circuit in Fig.~\ref{fig:circuit2p}
and $r = 2$ stochastically inserts a pair of CNOTs accompanying each of the three CNOTs either in the first or second half of the plaquette operator.
The combined noise and gate fidelity of the device were found to limit the ability to extrapolate further in CNOT noise, even with a single Trotter step.
It can be seen that these error mitigation techniques have allowed calculation of the electric energy associated with the SU(2) gauge field to the precision obtained after a single Trotter step.

It is important to determine the feasibility of retaining gauge-invariant Hilbert spaces with near-term quantum hardware.
For our calculations on IBM's {\tt Tokyo} quantum device,
before CNOT extrapolation, the $(N_{\text{Trot}},r) = (1,1)$ measurements were found to remain in the gauge invariant space with a survival population of $\sim45\%$, as shown in the bottom panel of Fig.~\ref{fig:IBMdata}.
After linear extrapolation in the probabilities, this was increased to $\sim65\%$,
with survival population decreasing as evolution time increases.
The survival population for $N_{\text{Trot}}=2$ was found to be $\sim25\%$,
consistent with loss of quantum coherence of a four-dimensional physical space embedded onto four qubits,
precluding CNOT extrapolation.
This observable is a diagnostic of the calculation.
As it approaches the decorrelated limit (25\%), CNOT extrapolations become less reliable leading to the underestimate of systematic uncertainties in Fig.~\ref{fig:IBMdata}.
Because neither the proposed qubit representation, nor the subsequent Trotterization, produce gauge-variant error contributions,
the observed decay of population in the physical subspace is a measure of the device's ability to robustly isolate Hilbert subspaces.
This is likely to be an essential capability for evolving lattice gauge theories and other systems with conserved quantities, as well as for quantum error correction.


When increasing $\Lambda_j$, the plaquette operator must be calculated and designed over 8 qubit registers, each containing $\lceil\log_2 \left( 2 \Lambda_j +1 \right)\rceil$ qubits.
The classical computational resources required to define this operator with Eq.~\eqref{eq:plaquettematrixelement} scales with the number of unique non-zero matrix elements, which is polynomial in $\Lambda_j$.
When constructing the time evolution operator for $\Lambda_j > 1/2$, the combination of Trotterization and Pauli decomposition of the 4-register operators in $j_{\ell, r}$-controlled sectors generically generates interactions breaking gauge invariance \cite{Zohar:2013zla,Stannigel:2013zka,Stryker:2018efp}.
The breaking of gauge invariance will be important to control if this decomposition is used in future calculations.


Developing quantum computation capabilities for non-Abelian gauge field theories is a major objective of nuclear physics and high-energy physics research.
One of the challenges facing such calculations is that the mapping of the gauge theory onto a discretized lattice involves a digitization of the gauge fields.
We have presented calculations of the dynamics of a one-dimensional SU(2) plaquette string with implementation on IBM's Q Experience superconducting hardware.
This was made feasible by an improved mapping of the angular momentum basis states describing link variables.
Our design of the plaquette operator for digital quantum devices requires local control from qubit registers beyond the active plaquette.
This key feature is expected to persist in future developments of quantum computing for gauge theories.
Extension of this analytic reduction beyond one dimension is naturally suited to lattices with three-point vertices.
Comparisons, at the level of explicit digital implementation,
of this mapping with proposed alternatives will be of importance for realizing physically-relevant quantum computations of non-Abelian gauge theories.

\vskip 0.2in
\begin{acknowledgments}
We would like to thank David Kaplan, Indrakshi Raychowdhury  and Erez Zohar for important discussions.
We would also like to thank Donny Greenberg, the IBM quantum computing research group, and the ORNL Q Hub for informative discussions and facilitating access to the IBM quantum devices.
We acknowledge use of the IBM Q for this work. The views expressed are those of the authors and do not reflect the official policy or position of IBM or the IBM Q team.
We thank ORNL collaborators Raphael Pooser and Pavel Lougovski with the U.S. Department of Energy, Office of Science, Office of Advanced
Scientific Computing Research (ASCR) quantum Testbed Pathfinder program and quantum algorithm teams program, under field work proposal numbers ERKJ335 and ERKJ333.
This work was performed, in part,
at the Aspen Center for Physics, which is supported by National Science Foundation grant PHY-1607611.
NK, JRS, and MJS were supported by the Institute for Nuclear Theory with DOE grant No. DE-FG02-00ER41132, Oak Ridge National Laboratory subcontract 4000158760, and Fermi National Accelerator Laboratory PO No. 652197.  NK was supported
in part by a Microsoft Research PhD Fellowship, and JRS was supported in part by the National Science Foundation Graduate Research Fellowship Program under Grant No. DGE-1256082.
\end{acknowledgments}

\bibliography{bibdnaq}
\onecolumngrid


\begin{center}
\large\textbf{Supplemental Material for\\ \enquote{Non-Abelian SU(2) gauge theory in one dimension on digital quantum computers}}
\end{center}

\section{Hamiltonians, Operators, and Data Tables}
For the two-plaquette lattice in one dimension with periodic boundary conditions implemented on the IBM quantum device {\tt Tokyo}, the Hamiltonian implemented in the full 16-dimensional Hilbert space with the chosen GVC is
\begin{equation}
\setcounter{MaxMatrixCols}{16}
  \mathcal{H}^{(1/2)} = \frac{1}{2g^2} \scalemath{0.9}{
  \begin{pmatrix}
 \cellcolor{green!10}
 0 & 0 & 0 & 0 & 0 & 0 & 0 & \cellcolor{green!10} -2 & 0 & 0 & \cellcolor{green!10} 0 & 0 & 0 & \cellcolor{green!10}-2 & 0 & 0 \\
 0 & \frac{3 g^4}{4} & 0 & 0 & 0 & 0 & -2 & 0 & 0 & 0 & 0 & 0 & -2 & 0 & 0 & 0 \\
 0 & 0 & \frac{3 g^4}{2} & 0 & 0 & -2 & 0 & 0 & 0 & 0 & 0 & 0 & 0 & 0 & 0 & -\frac{1}{2} \\
 0 & 0 & 0 & \frac{9 g^4}{4} & -2 & 0 & 0 & 0 & 0 & 0 & 0 & 0 & 0 & 0 & -\frac{1}{2} & 0 \\
 0 & 0 & 0 & -2 & \frac{3 g^4}{4} & 0 & 0 & 0 & 0 & -2 & 0 & 0 & 0 & 0 & 0 & 0 \\
 0 & 0 & -2 & 0 & 0 & \frac{3 g^4}{2} & 0 & 0 & -2 & 0 & 0 & 0 & 0 & 0 & 0 & 0 \\
 0 & -2 & 0 & 0 & 0 & 0 & \frac{9 g^4}{4} & 0 & 0 & 0 & 0 & -\frac{1}{2} & 0 & 0 & 0 & 0 \\
 \cellcolor{green!10} -2 & 0 & 0 & 0 & 0 & 0 & 0 & \cellcolor{green!10}3 g^4 & 0 & 0 & \cellcolor{green!10}-\frac{1}{2} & 0 & 0 & \cellcolor{green!10}0 & 0 & 0 \\
 0 & 0 & 0 & 0 & 0 & -2 & 0 & 0 & \frac{3 g^4}{2} & 0 & 0 & 0 & 0 & 0 & 0 & -\frac{1}{2} \\
 0 & 0 & 0 & 0 & -2 & 0 & 0 & 0 & 0 & \frac{9 g^4}{4} & 0 & 0 & 0 & 0 & -\frac{1}{2} & 0 \\
 \cellcolor{green!10}0 & 0 & 0 & 0 & 0 & 0 & 0 & \cellcolor{green!10}-\frac{1}{2} & 0 & 0 & \cellcolor{green!10} 3 g^4 & 0 & 0 & \cellcolor{green!10}-\frac{1}{2} & 0 & 0 \\
 0 & 0 & 0 & 0 & 0 & 0 & -\frac{1}{2} & 0 & 0 & 0 & 0 & \frac{15 g^4}{4} & -\frac{1}{2} & 0 & 0 & 0 \\
 0 & -2 & 0 & 0 & 0 & 0 & 0 & 0 & 0 & 0 & 0 & -\frac{1}{2} & \frac{9 g^4}{4} & 0 & 0 & 0 \\
 \cellcolor{green!10}-2 & 0 & 0 & 0 & 0 & 0 & 0 & \cellcolor{green!10}0 & 0 & 0 & \cellcolor{green!10}-\frac{1}{2} & 0 & 0 & \cellcolor{green!10} 3 g^4 & 0 & 0 \\
 0 & 0 & 0 & -\frac{1}{2} & 0 & 0 & 0 & 0 & 0 & -\frac{1}{2} & 0 & 0 & 0 & 0 & \frac{15 g^4}{4} & 0 \\
 0 & 0 & -\frac{1}{2} & 0 & 0 & 0 & 0 & 0 & -\frac{1}{2} & 0 & 0 & 0 & 0 & 0 & 0 & \frac{9 g^4}{2}
  \end{pmatrix} }
\end{equation}
with matrix elements of the four-dimensional physical subspace highlighted.
For the chosen coupling of $g^2 = 0.2$, the ground state energy density per plaquette, through exact diagonalization, is calculated to be -3.5658 and the lowest energy gap (the observable associated with the \enquote{SU(2)-glueball} mass in the infinite volume limit) is calculated to be 7.4139.  The structure of the ground state wavefunction is
\begin{equation}
  |\psi_{gs}\rangle = 0.6943 \begin{gathered}\includegraphics[width = 0.08\textwidth]{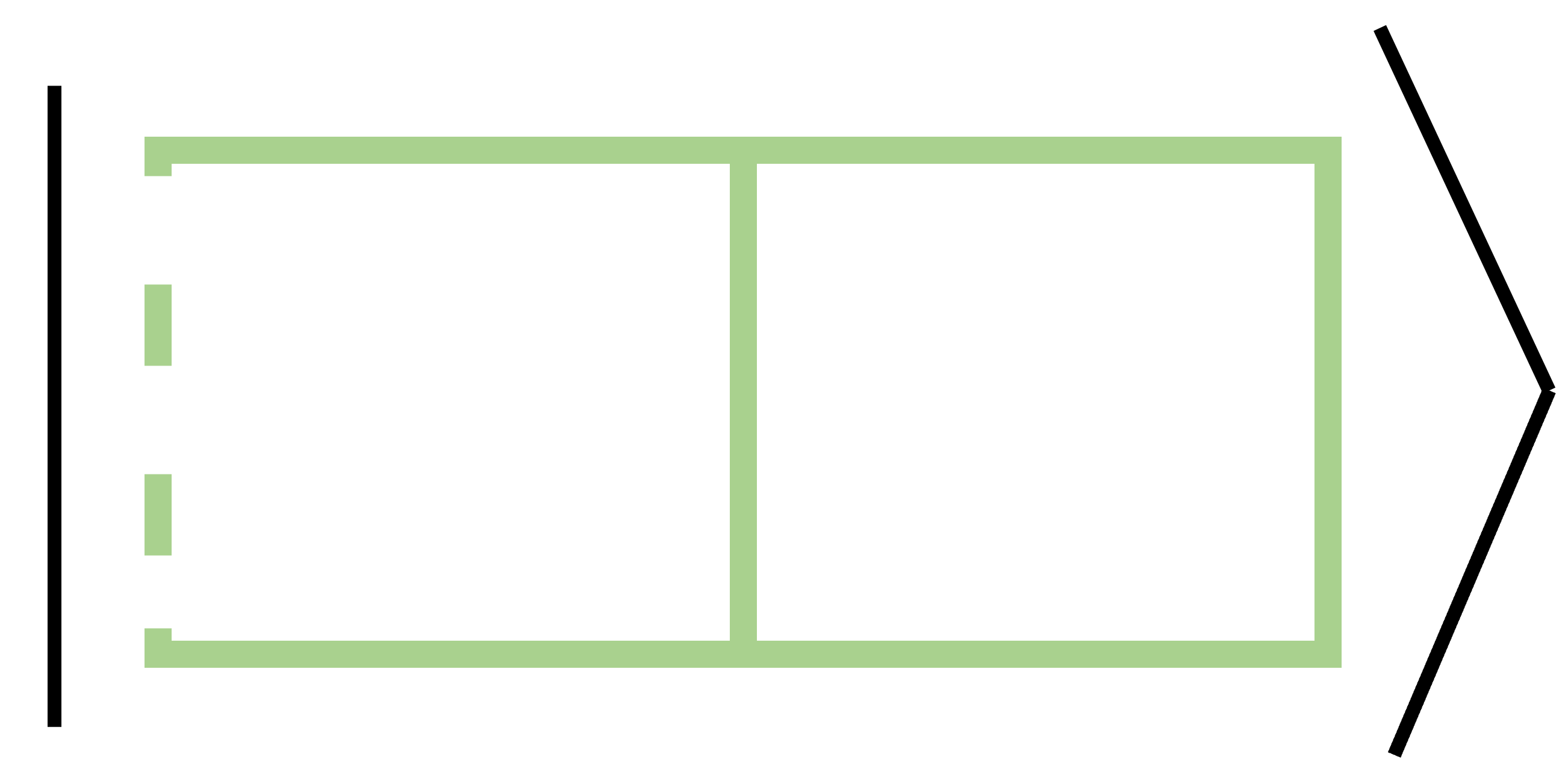} \end{gathered} +  0.1666 \begin{gathered}
    \includegraphics[width = 0.08\textwidth]{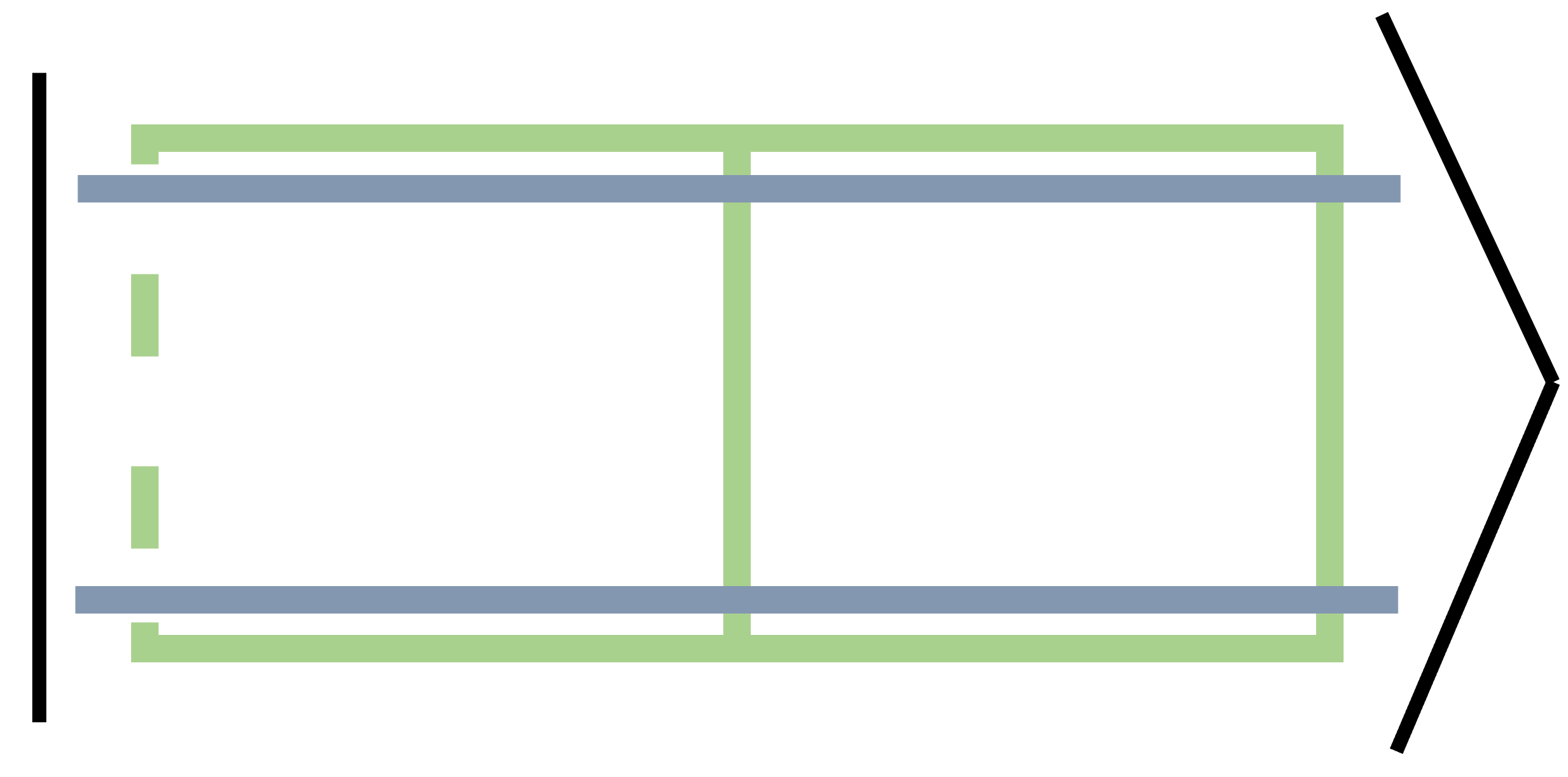}
  \end{gathered} + 0.4951 \left( \begin{gathered}\includegraphics[width = 0.08\textwidth]{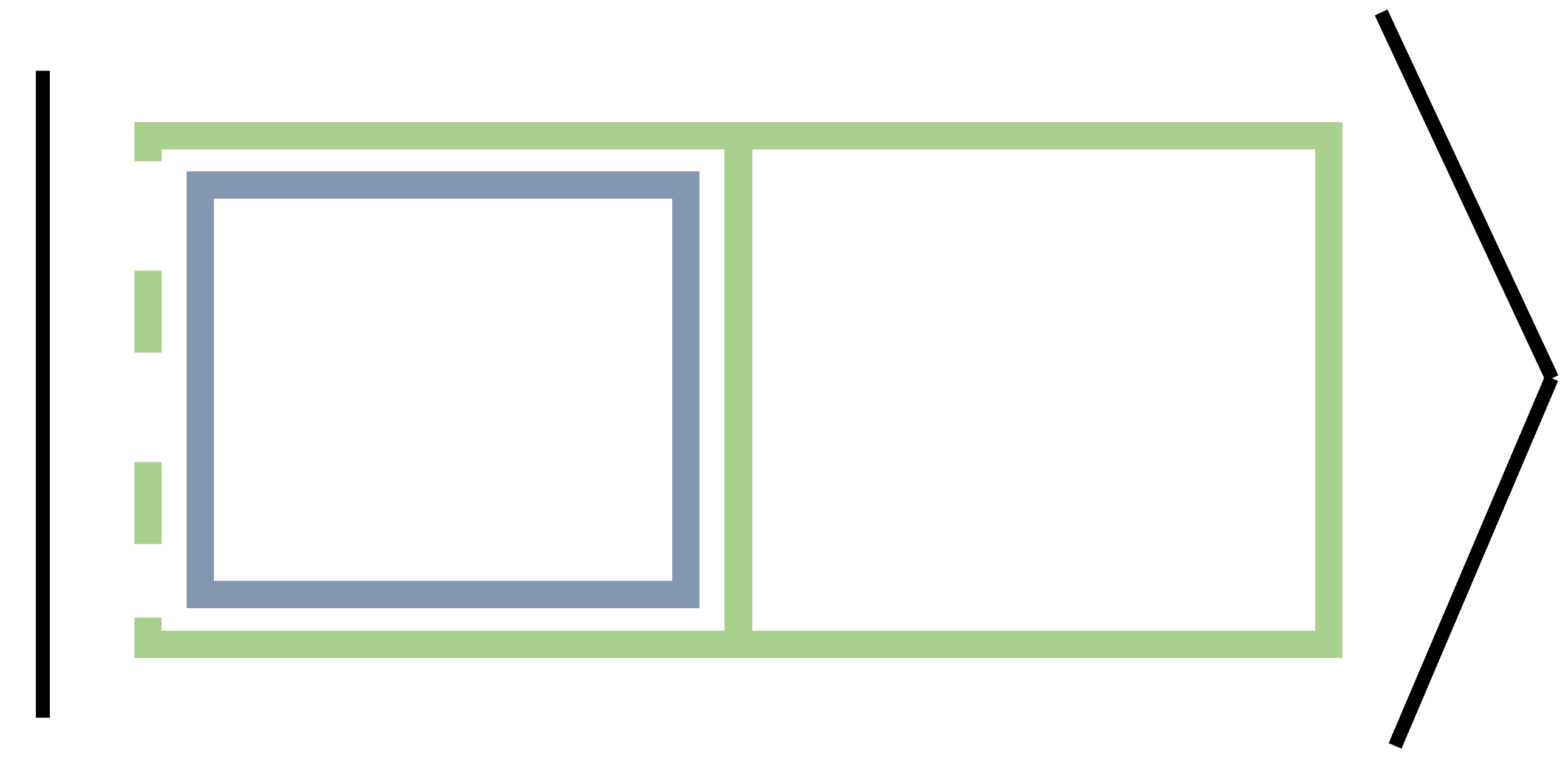} \end{gathered} + \begin{gathered}\includegraphics[width = 0.08\textwidth]{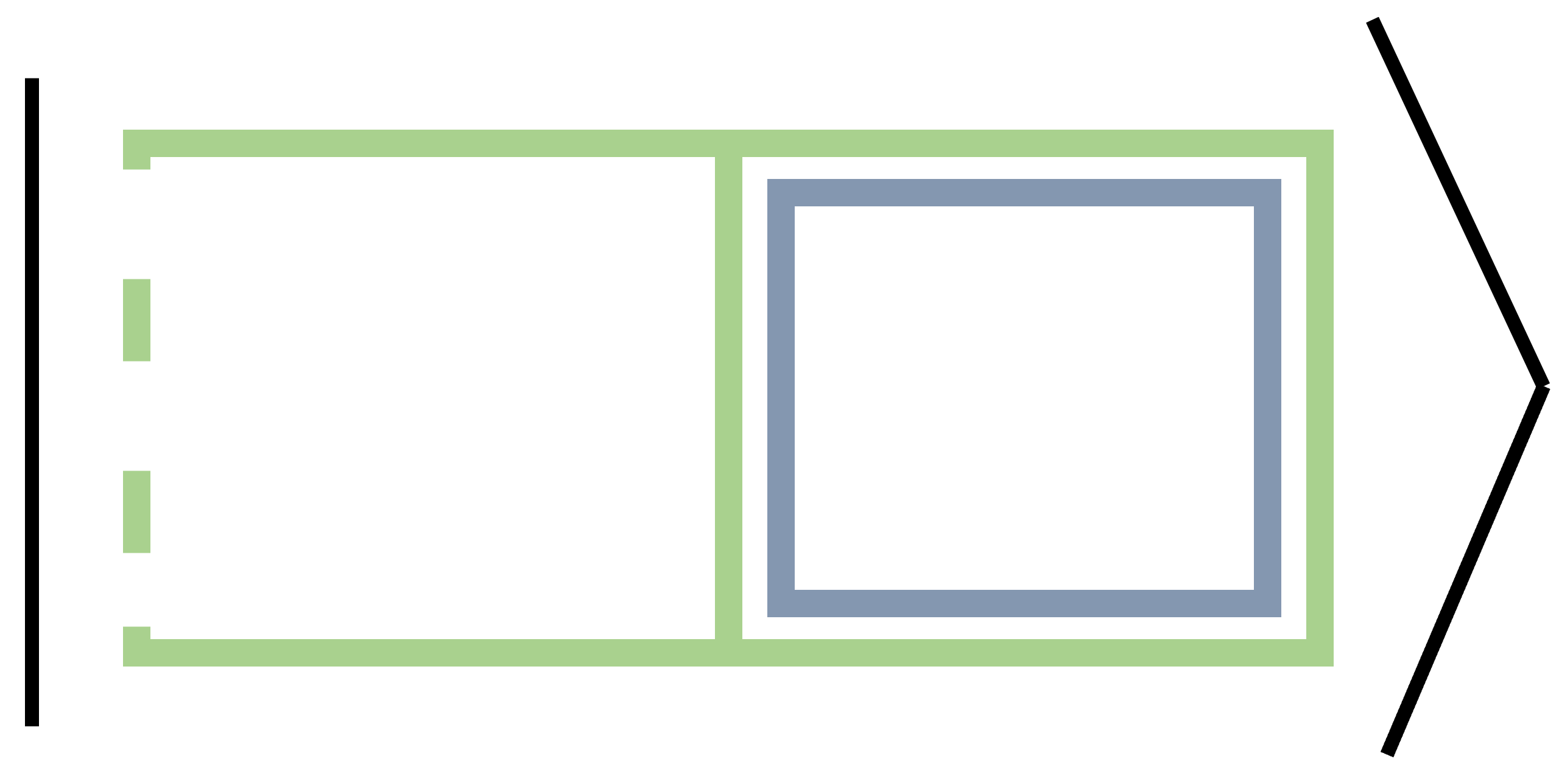} \end{gathered} \right) \ \ \ .
\end{equation}
On each link, a single line corresponds to $j = 0$ and a double line corresponds to $j = 1/2$.
The first electric, single plaquette operator in the full 16-dimensional space is diagonal
\begin{equation}
  E^2_{\Box_1} = \frac{g^2}{2}\text{diagonal}\left( 0, \frac{3}{4}, 0, \frac{3}{4}, \frac{3}{4}, \frac{3}{2}, \frac{3}{4}, \frac{3}{2}, \frac{3}{2}, \frac{9}{4}, \frac{3}{2}, \frac{9}{4}, \frac{9}{4}, 3, \frac{9}{4}, 3 \right)
\end{equation}
with matrix elements serving as weights of the measured probabilities in the measurement of the electric energy expectation value as shown in Fig.~\ref{fig:IBMdata}.  The data points in Fig.~\ref{fig:IBMdata} are presented in Tables~\ref{tab:electricenergy} and~\ref{tab:survivalprob}.
\begin{table}[h]
  \centering
  $N_{\text{Trot}} = 1$ \hspace{2cm}   $N_{\text{Trot}} = 2$\\
  \begin{tabular}{cc}
    \hline
    \hline
    time & $\langle H_{E,\Box_1}\rangle $ \\
    \hline
    0.02&0.009(9)\\
    0.07&0.052(6)\\
    0.12&0.127(7)\\
    0.17&0.201(12)\\
    0.22&0.261(10)\\
    0.27&0.282(7)\\
    0.32&0.278(8)\\
    0.37&0.254(6)\\
    \hline
    \hline
  \end{tabular}
    \hspace{1cm}
  \begin{tabular}{cc}
    \hline
    \hline
    time & $\langle H_{E,\Box_1}\rangle $ \\
    \hline
    0.02&0.027(14)\\
    0.07&0.074(14)\\
    0.12&0.124(14)\\
    0.17&0.159(10)\\
    0.22&0.186(15)\\
    0.27&0.177(12)\\
    0.32&0.144(20)\\
    0.37&0.093(18)\\
    \hline
    \hline
  \end{tabular}
  \caption{Numerical values of the expectation value of the single electric plaquette energy contribution for time evolutions implemented with 1,2 Trotter steps as measured on IBM's quantum device {\tt Tokyo}~\cite{ibm} shown in the top panel of Fig.~\ref{fig:IBMdata}. Uncertainties represent statistical variation, as well as a systematic uncertainty estimated from reproducibility measurements.}
  \label{tab:electricenergy}
  \vspace{1cm}
   \hspace{1cm} (1,1) \hspace{2.5cm} (1,2) \hspace{2.4cm} (2,1) \hspace{2.3cm} (2,2) \hspace{1.3cm} Linear Extrapolation \\
   \centering
    \begin{tabular}{cc}
    \hline
    \hline
    time & Survival Prob. \\
    \hline
    0.02&0.47(1)\\
    0.07&0.49(2)\\
    0.12&0.48(2)\\
    0.17&0.47(2)\\
    0.22&0.43(1)\\
    0.27&0.41(2)\\
    0.32&0.39(1)\\
    0.37&0.37(1)\\
    \hline
    \hline
    \end{tabular}
    \hspace{0.1cm}
    \begin{tabular}{cc}
    \hline
    \hline
    time & Survival Prob. \\
    \hline
    0.02&0.29(2)\\
    0.07&0.31(3)\\
    0.12&0.28(2)\\
    0.17&0.27(1)\\
    0.22&0.25(1)\\
    0.27&0.25(2)\\
    0.32&0.23(1)\\
    0.37&0.21(1)\\
    \hline
    \hline
    \end{tabular}
    \hspace{0.1cm}
    \begin{tabular}{cc}
    \hline
    \hline
    time & Survival Prob. \\
    \hline
    0.02&0.23(1)\\
    0.07&0.24(2)\\
    0.12&0.26(2)\\
    0.17&0.24(1)\\
    0.22&0.25(2)\\
    0.27&0.23(2)\\
    0.32&0.22(1)\\
    0.37&0.22(1)\\
    \hline
    \hline
    \end{tabular}
    \hspace{0.1cm}
    \begin{tabular}{cc}
    \hline
    \hline
    time & Survival Prob. \\
    \hline
    0.02&0.27(1)\\
    0.07&0.27(1)\\
    0.12&0.24(1)\\
    0.17&0.26(1)\\
    0.22&0.25(2)\\
    0.27&0.26(1)\\
    0.32&0.28(1)\\
    0.37&0.26(1)\\
    \hline
    \hline
    \end{tabular}
    \hspace{0.1cm}
    \begin{tabular}{cc}
    \hline
    \hline
    time & Survival Prob. \\
    \hline
    0.02&0.630(14)\\
    0.07&0.640(16)\\
    0.12&0.659(25)\\
    0.17&0.647(33)\\
    0.22&0.572(17)\\
    0.27&0.554(14)\\
    0.32&0.535(17)\\
    0.37&0.527(17)\\
    \hline
    \hline
  \end{tabular}
  \caption{Survival probabilities in the physical subspace as measured on IBM's quantum device {\tt Tokyo}~\cite{ibm} shown in the bottom panel of Fig.~\ref{fig:IBMdata}.  The label indicates $\left(N_{\text{Trot}}, r\right)$ values. The linear extrapolation is determined by extrapolation of computational basis state probabilities in $r$ for $N_{\text{Trot}} = 1$.  Uncertainties represent statistical variation, as well as a systematic uncertainty estimated from reproducibility measurements.}
  \label{tab:survivalprob}
\end{table}

\vfill

\end{document}